# Controlling Ferromagnetic Ground States and Solitons in Thin Films and Nanowires built from Iron Phthalocyanine Chains

Z. Wu[1, 2, †], P. Robaschik[1, 2], L. R. Fleet[1, 2], S. Felton[1, 2, ‡], G. Aeppli[3, 4, 5] and S. Heutz[1, 2]*

[1] *Department of Materials, Imperial College London, London, SW7 2AZ, UK.*
[2] *London Centre for Nanotechnology, Imperial College London, London, SW7 2AZ, UK.*
[3] *Department of Physics, ETH, 8093 Zürich, Switzerland*
[4] *Institut de Physique, EPFL, 1015 Lausanne, Switzerland*
[5] *Paul Scherrer Institut, 5232 Villigen, Switzerland*

† Current address: School of Physics and Optoelectronic Engineering and Instrumentation Science, Dalian University of Technology, Dalian 116024, China.

‡ Current address: School of Mathematics and Physics, Queen's University Belfast, Belfast BT7 1NN, UK.

* e-mail: s.heutz@imperial.ac.uk

**Iron phthalocyanine (FePc) is a molecular semiconductor whose building blocks are one-dimensional ferromagnetic chains. We show that its optical and magnetic properties are controlled by the growth strategy, obtaining extremely high coercivities of over 1 T and modulating the exchange constant between 15 and 29 K through tuning the crystal phase by switching from organic molecular beam deposition, producing continuous thin films of nanocrystals with controlled orientations, to organic vapour phase deposition, producing ultralong nanowires. Magnetisation measurements are analysed using a suite of concepts and simply stated formulas with broad applicability to all one-dimensional ferromagnetic chains. They show that FePc is best described by a Heisenberg model with a preference for the moments to lie in the molecular planes, where the chain Hamiltonian is very similar to that for the classic inorganic magnet $CsNiF_3$, but with ferromagnetic rather than antiferromagnetic interchain interactions. The data at large magnetic fields are well-described by the soliton picture, where the dominant (and topologically non-trivial) degrees of freedom are moving one-dimensional magnetic domain walls, which was successful for $CsNiF_3$, and at low temperatures and fields by the "super-Curie-Weiss" law of $1/(T^2+\theta^2)$ characteristic of nearly one-dimensional xy and Heisenberg ferromagnets. The ability to control the molecular orientation and ferromagnetism of FePc systems, and produce them on flexible substrates as thin films or nanowires, taken together with excellent transistor characteristics reported previously for nanowires of copper and cobalt analogues, makes them potentially useful for magneto-optical and spintronic devices.**

## Introduction

Organic spintronics is rapidly developing with many reports on the physics of potential host systems, and classical organic spin-valves as well as spin-based quantum computers are considered promising for future applications.[1, 2] Despite recent progress, still far from well-understood are spin-injection and spin-transport in organic semiconductors.[3-7] Difficulties arise because currently the sources of spin-polarised carriers are conventional ferromagnets, such as permalloy, or oxide ferromagnets such as lanthanum strontium manganite.[7, 8] Although the interfaces give rise to interesting new phenomena due to doping or hybridisation,[9-11] their ill-defined character makes the science of the resulting spin valves difficult. It would be preferable to use organic layers for injection, transport and detection as here there would be mechanically and chemically compatible interfaces suited to use with flexible substrates. Currently, however, there are very few molecular magnets stable at high temperature, with the exception of the V(TCNE)$_x$ family of materials,[12] which have been associated with delicate processing and stability,[13-15] although recent progress on encapsulation could help the integration for applications.[16]

Transition metal phthalocyanines (MPcs) are very interesting as a new type of organic magnetic semiconductor that can offer tunability of the magnetic properties.[17-21] Pcs are mostly planar aromatic molecules, existing in a large number of crystal phases, which can host a spin-bearing ion in the centre. As the magnetic properties are strongly dependent on the metal ion and orbital overlap, these molecules can be combined to create heterostructures for spintronics. We have recently shown that antiferromagnetic couplings of up to 100 K can be obtained in cobalt phthalocyanine (CoPc) films, and theory suggests that this could be further enhanced to above room temperature through further tuning of the structure.[21] In addition, CoPc (as well as CuPc) nanowires, with no selection or post-processing, have recently yielded high performance transistors.[22] The next step is to search for ferromagnetic analogues which are more susceptible to small external magnetic fields and for which electron gases in transport channels (e.g. between source and drain in a transistor) could be spin polarised. Iron phthalocyanine (FePc) has already been shown to exhibit ferromagnetism[20] at low temperature with some tuneability of the magnetic properties achievable through altering the crystal structure[23] or grain size,[24, 25] and is characterised by an unusually high orbital moment.[26] Ferromagnetic FePc nanowires with crystal structure identical to the thin films were recently highlighted, with patterned substrate and inorganic shells promoting vertical growth.[25] However, the magnetic anisotropy, as deduced from differences in coercivity for hysteresis curves measured along perpendicular directions, was opposite to literature values obtained from films and more conclusive data are required. Finally, we have demonstrated an FePc transistor based on conventional thin film deposition.[22] Analogous to the other metal Pcs, FePc consists of quasi one-dimensional (1D) FePc chains along the stacking axis (b-axis), resulting in control of the molecular overlap within chains *via* creation of an appropriate stacking polymorph.

In this report we use 3,4,9,10-perylenetetracarboxylic dianhydride (PTCDA) to change the orientation of the FePc molecules from nearly perpendicular to the substrate surface to lying almost parallel to it, meaning that the direction of the 1D magnetic chains can be modified easily on many substrates. This system is also compatible with transparent electrodes which will enable electro-magneto-optical experiments and applications.[27] Furthermore, we report the structural and magnetic properties of FePc nanowires fabricated using organic vapour phase deposition (OVPD). The low dimensionality of the nanowires, which themselves consist of 1D ferromagnetic chains, leads to a coercivity of over 1 T at 2 K. This is more than one order of magnitude larger than values reported for FePc thin films and could be further enhanced through additional optimisation of the stacking angles. We introduce a description of FePc as a 1D xy chain containing solitons, not only enabling ready estimation of interchain couplings and the magnetic anisotropy, but also representing topological excitations which could interact with charge carriers in future FePc wire-based transistors. We find that the anisotropy energy is 62 K for the wires, substantially higher than for the films ($D$ = 33 K), which correlates with the high coercivity and different structure.

**Experimental Methods**

FePc films were grown by conventional organic molecular beam deposition (OMBD), using a commercial SPECTROS system from Kurt J. Lesker. Fe(II)Pc powder (98% purity), purchased from Sigma-Aldrich, was purified twice using conventional gradient sublimation in a vacuum of ~ 2 x $10^{-2}$ mbar and a nitrogen carrier gas. The 100 nm FePc films were grown onto Kapton (polyimide), silicon and glass. For the templated samples, a 20 nm thick PTCDA layer was also deposited onto Kapton to induce structural templating, followed by a 100 nm thick FePc layer, without breaking the vacuum. Both templated and non-templated samples were deposited at the same time. For the depositions, we used an effusion cell, with a chamber base pressure of 5 x $10^{-7}$ mbar, and the substrate maintained at room temperature. The substrates were rotated during the deposition to ensure film uniformity. Deposition rates were 0.5 Å/s and 0.1 Å/s for the FePc and PTCDA growths, respectively.

The FePc nanowires were deposited by means of organic vapour phase deposition (OVPD), using a three-zone TMH12/75/750 Elite furnace with individual temperature controls for each zone. The OVPD chamber consists of a quartz tube (2 m long and 4 cm in diameter) inserted into the furnace where the temperature of the individual zones (each ~ 33 cm in length) can be independently controlled. We placed the purified FePc crystals in a crucible in the sublimation zone, set to 480 °C. Nitrogen was the carrier gas, with a rate of 1.0 l/min, to transport the evaporated molecules along the quartz tube to the cooler furnace zone, with the temperatures in the second and third zones 500 °C and 250 °C, respectively. The sharp temperature gradient causes quick condensation of the molecules and nucleation, and subsequent growth, of FePc nanowires. The growth duration was approximately 48 hours. A large piece of Kapton 25 μm thick, 125 mm in width and 800 mm in length was used to cover the wall of the inner tube and provide the substrate onto which the nanowires grow, and wires were produced across the whole 80 cm length of the kapton. Additional substrates were also placed inside the furnace for structural and magnetic characterisation, discussed below.

We employed a SQUID-based magnetic property measurement system (MPMS-7) from Quantum Design for films and wires grown on Kapton. Through *in-situ* shadow masking in the OMBD and OVPD setups, a rectangular strip of molecular films or wires 3 mm wide and 70 mm long was produced, with bare substrate about 40 mm wide and 70 mm long on either side of the strip. The 70 mm x 83 mm sample was then rolled into the sample straw. This geometry eliminates the background signal from the Kapton substrate, as each side of the substrate is longer than the measuring coils, thus cancelling the background signal, as described elsewhere.[19] We have identified a magnetic impurity corresponding to 0.014 $\mu_B$ (i.e. 1% of the total signal) in the templated film, and this has been subtracted using the method described in the supplementary information. We ascribe this impurity with the higher sticking coefficient of PTCDA compared to kapton, which might promote the adsorption of impurities, as has been noted before for $F_{16}CuPc$.[28]

A LEO Gemini 1525 field emission gun scanning electron microscope (SEM) with 5 kV accelerating voltage, yielded morphologies of samples coated with a thin (10-15 nm) chromium layer. The transmission electron microscopes (TEM) used were JEOL 2010 and JEOL 200FX instruments operated at 200 kV. The X-ray diffraction (XRD) instrument was a Panalytical X-pert operated in the $\theta$-$2\theta$ mode using Cu K$_\alpha$ radiation (40 kV, 40 mA, nickel filter) and a step size of $2\theta$ = 0.033° with a counting time of 80 s per step. A Perkin-Elmer Lambda 25 UV-Vis spectrometer provided electronic absorption spectra. For the TEM and transmission electron diffraction (TED) analysis, we placed holey carbon copper grids (Agar) on the Kapton in the chamber during the growth. The images were obtained for nanowires oriented with a low index crystal zone axis parallel to the electron beam. SEM and XRD were performed on wires collected from Kapton, glass and silicon substrates, all obtained *in situ*.

It should be noted that no post-deposition annealing or processing was performed on any of the samples.

**Results and discussion**

To characterise the highly anisotropic FePc magnetism, we prepare FePc films with different orientations with respect to the substrate, using the templating technique that was previously applied to other planar phthalocyanines.[29, 30] The unique orientation of the molecules with respect to the magnetic field enables us to model the magnetic properties using the Heisenberg model with planar anisotropy and to extract both the magnetic coupling strength *J* and the anisotropy constant *D*.

We then investigate FePc nanowires, which represent a particularly interesting system as the chain length can be up to five orders of magnitude larger than for the films,[31] a configuration that significantly improves the electronic transport properties.[22] It has already been shown that the length of the iron chains also strongly influences the magnetic properties, with an increase in coercive field observed for increasing crystal size.[32] Therefore new behaviours are expected for FePc nanowires with the magnetic anisotropy of the molecular plane competing with the shape anisotropy of the nanowires.

*Controlling spin chain orientation in FePc films*

SEM and TEM show that both the non-templated and templated films consist of small spherical grains, approximately 30 to 40 nm in diameter, uniformly covering the substrate, see Figure 1. This morphology is typical for thin films of planar Pcs deposited at room temperature by OMBD.[24, 33] Using TEM the differences in the film structure start to become clear. From the digital diffractograms, and TEM lattice fringes, the plane spacings observed in the non-templated (Figure 1c) and templated (Figure 1d) films are 12.1 and 12.8 Å, respectively. These planes are normal to the substrate and correspond to the (001) and (100) spacings respectively when using the $\alpha$-phase structure derived by Hoshino *et al.* for CuPc.[34] We note that the existing indexed structures for FePc correspond to a high temperature preparation traditionally associated with the β-phase,[35-37] and that using the structure of α-CuPc to interpret the diffraction features of FePc films deposited at room temperature is an approach that has been commonly used by other groups.[20, 23, 38] The TEM pattern is as expected from previous studies of the effect of PTCDA on metal phthalocyanine (MPc) films,[34, 39-41] although it has been observed that the templated phase can also display (001) planes in TEM due to the spread in the orientation of the molecular crystals.[30, 39]

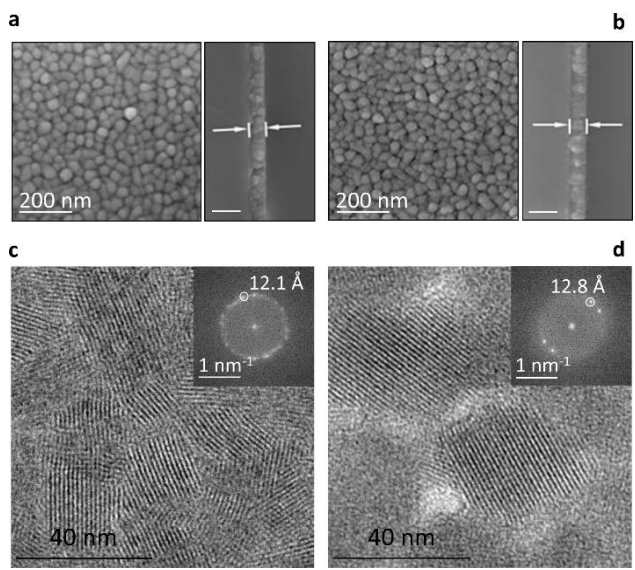

**Figure 1. Templating FePc thin films.** SEM images of **a**, FePc on silicon and **b**, templated FePc, with their corresponding cross-sections. TEM images of **c**, FePc and **d**, templated FePc show similar morphologies and a high degree of crystallinity within the grains (inset, diffraction pattern of the corresponding films).

The orientation of the films can also be verified using XRD. An intense peak at $2\theta = 6.9°$ was observed in the XRD patterns, Figure 2a-b, for the non-templated films, corresponding to diffraction from the (100) plane of α-phase Pc.[34] The absence of other peaks suggests that the film is preferentially grown with the (100) plane parallel to the substrate, agreeing with the TEM observations of the (001) perpendicular planes. Another weak peak at $2\theta = 13.8°$ arises from the (200) plane, the second harmonic peak. The angle between the (100) crystallographic plane, parallel to the substrate, and the molecular plane is 82°, with the stacking axis parallel to the substrate. For the templated film, two different peaks were observed at $2\theta = 26.7°$ and 27.8°, Figure 2a and c. These correspond to diffraction from the (01-2) and (11-2) planes of the α-polymorph respectively.[34] The angles between the molecular and crystallographic planes of (01-2) and (11-2) are 9.0° and 7.5°, respectively. Introducing a PTCDA templating layer therefore causes the molecules to lie approximately parallel to the substrate. Although both (01-2) and (11-2) diffraction peaks were observed, their texture coefficients[42], as deduced from the relative intensity of peaks identified in the $2\theta = 25\text{-}30°$ region, are 0.3 and 1.7 respectively. The preponderance of the (11-2) peak implies that the templated film is preferentially oriented along this direction, which may be because this molecular arrangement corresponds to an orientation more parallel to the PTCDA. The molecular orientations in both films are summarised in Figure 2d and e.

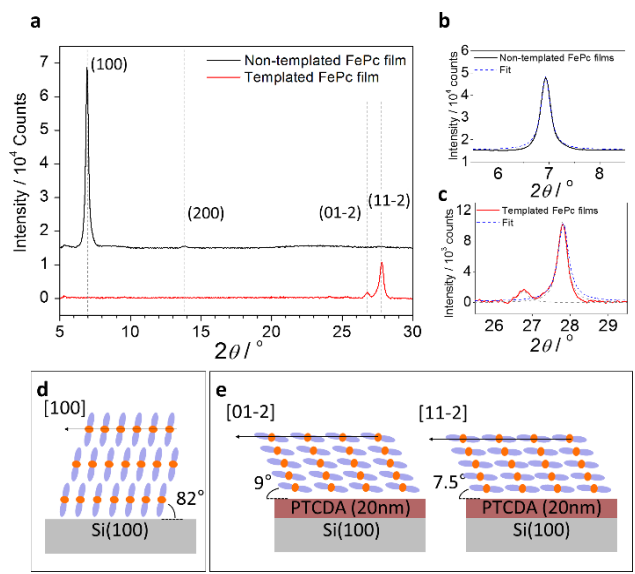

**Figure 2. Structure of FePc thin films. a**, XRD scans and Lorentzian fits for **b**, non-templated and **c**, templated FePc 100 nm thick films. The corresponding molecular orientation is shown for **d**, non-templated and **e**, templated films.

The average crystallite sizes were calculated by using a Lorentzian fit to the diffraction peaks at $2\theta = 6.8°$ of the non-templated and $2\theta = 26.7°$ and $27.8°$ of the templated films (Figure 2b and c respectively), with the full width at half maximum (FWHM) related to the grain size through the Scherrer equation.[43] The instrumental broadening of ~ 0.13° does not significantly affect the result. The average grain size of the non-templated film was found to be 41 ± 8 nm. For the templated film the average grain sizes were found to be 30±6 and 32±6 nm, respectively. Thus, the estimated grain sizes of both non-templated and templated films are similar, and, in agreement with the estimates from SEM and TEM in Figure 1.

*FePc nanowires*

Nanowires of Pcs with high aspect ratio can be obtained using specific conditions in OVPD and were first discovered for CuPc.[31] They were also obtained for CoPc,[22, 44] and recent integration into field effect transistors has highlighted their exceptionally high on-off ratio, competitive mobility and long lifetimes.[22] FePc nanowires were also recently produced in a high vacuum system with high substrate temperature and argon partial pressure up to 0.02 mbar, but this resulted in shorter lengths, and has not been attributed to the η-phase.[25]

Here we demonstrate that FePc can also form ultralong nanowires using OVPD (see Methods) which are at least several microns in length and approximately 40-100 nm in width, corresponding to extremely high aspect ratios, Figure 3a and b. It is also observed that the wire-like crystallites were oriented randomly and uniformly with respect to the surface of the substrate. As the optical properties of MPcs are closely related to the relative orientation and separation of adjacent molecules they are a useful tool for identifying the crystal structure. Two broad peaks, located at 635 and 800 nm, can be seen in the absorption spectrum of the FePc nanowires (Figure 3e), differing from the signature in the common phases of FePc, in particular the α-phase thin film grown by OMBD displayed as a comparison. The broadening of absorption for FePc corresponds to a higher hopping integral for the carriers along the chains, as was also found for CuPc. As the OVPD growth conditions used are known to form the η-polymorph for CuPc,[22, 31] it could be expected that the FePc nanowires crystallise in the same phase.

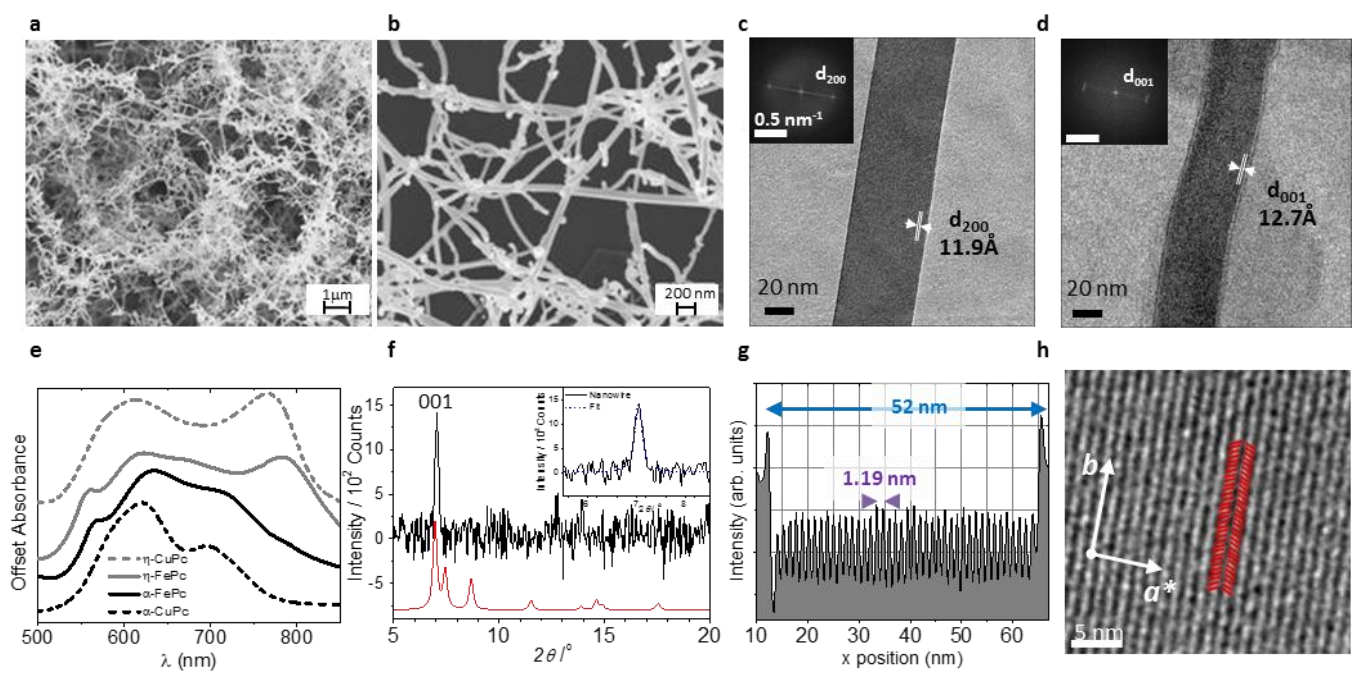

**Figure 3. Structure and morphology of FePc nanowires. a,b,** SEM images of FePc nanowire network. TEM images and digital diffractograms, shown in the insets, of single wires obtained in the **c**, (200) and **d**, (001) projections. **e**, normalised UV-Vis absorption of η- and α-FePc (continuous lines), with the equivalent CuPc spectra (dashed lines) shown for comparison and **f**, XRD scan (acquired for 80 s/step) for nanowires together with a simulated scan with a peak broadening of $2\theta = 0.2°$ in red.[31] **g**, linescan of the intensity for the wire shown in (c), highlighting the (200) lattice spacing and **h**, a HRTEM image with expected molecular arrangement.

The crystallinity and structure of the nanowires were further explored using TEM and TED, Figure 3c and d, and lattice spacings were extracted using ImageJ, see Figure 3g for a characteristic linescan for the (200) planes. The wires are highly crystalline, and the width was found to be approximately 50 nm, with lengths of at least several microns. Two common lattice plane spacings could be seen, at $d$ = 12.7 and 11.9 Å corresponding well with the (001) and (200) planes of the η-phase respectively, within error. In the η-phase the molecular plane lies at approximately 65° to the long wire axis in a herringbone arrangement represented on the high resolution TEM image in Figure 3h.

Using XRD a peak can be seen at 7.0°, agreeing well with the (001) diffraction peak of the η-polymorph (Figure 3f). The weak intensity of the peak and poor signal-to-noise ratio is most likely due to the small amount of FePc nanowires deposited onto the substrate and may also be due to the random distribution of orientations, observed using SEM, as opposed to the films which are highly textured and therefore have a higher contribution of crystallites in a favourable orientation for diffraction. The absence of other significant peaks in the XRD pattern is consistent with the high noise and large peak width. This obscures the other peaks which have a significantly lower structure factor than the (001) peak as seen by comparison with the simulated diffraction pattern. It is also possible that the nanowires align preferentially with the (001) crystallographic plane parallel to the substrate, although the (200) orientation parallel to the substrate can also be inferred from the TEM images. A Lorentzian fit was applied to the diffraction peak with the net FWHM of 0.20° corresponding to a length scale of 45 ± 9 nm, indistinguishable from the wire widths measured using electron microscopy.

*Magnetic properties of FePc films and nanowires*

As the PTCDA layer changes the direction of the stacking axis, it also changes the orientation of the 1D ferromagnetic chains. This was explored using SQUID magnetometry, Figure 4. Below 20 K the magnetisation increases rapidly to approximately 0.8 and 1.3 $\mu_B$/Fe for the non-templated (a and b) and templated (d and e) films, respectively. This indicates that the spins in both types of films have ferromagnetic correlations at $T < 20$ K. It can also be seen that the overall magnetisation of the templated film is larger than that of the non-templated film, with the magnetisation curve for the templated film appearing to be flatter and therefore closer to saturation in the high-field region. These differences arise due to magnetic anisotropy, with the moment preferring to lie in the plane of the molecule, as previously observed on sapphire and gold substrates, and depicted in Figure 4c.[20, 45, 46]

Hysteresis could be observed at 2 K with coercive fields of 80 and 35 mT obtained for the non-templated and templated films, respectively (Figure 4b and e, inset). Previous reports have indicated that the hysteresis in α-FePc powders is time-dependent.[23] The difference in coercive fields for systems with identical structure but different textures could therefore also be attributed to a slower relaxation when the field is not applied along the easy plane of magnetisation. An increase in coercivity (from 3 mT to 10 mT at 5 K) was also observed for short FePc standing nanowires when the applied field was rotated from parallel to perpendicular to the substrate, i.e. to a direction with a shorter projection along the molecular plane, which we define as the easy plane.[25]

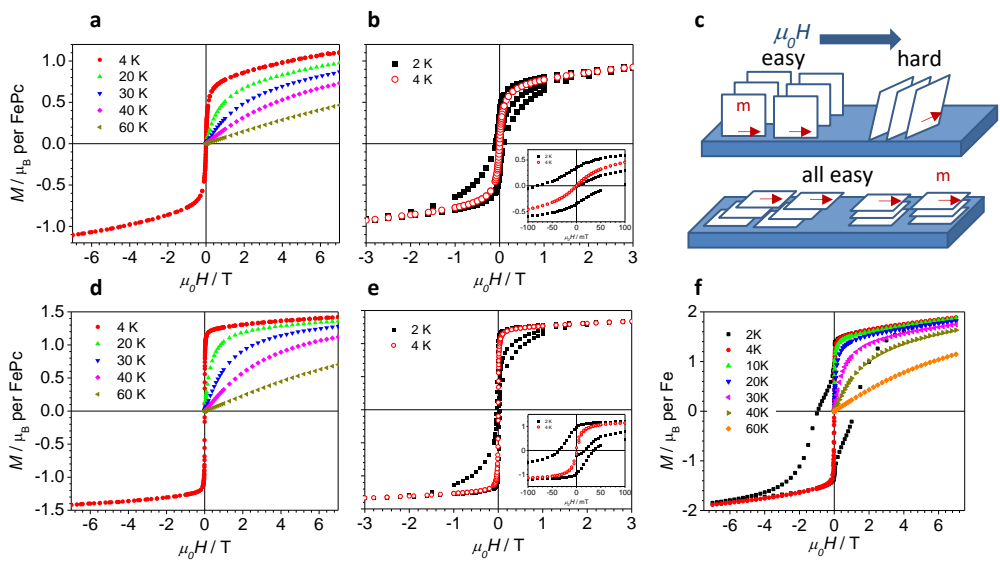

**Figure 4. Magnetic properties of FePc thin films.** Field-dependent magnetisation curves obtained at various temperatures for **a,** 100 nm non-templated and **d,** 100 nm templated FePc films with Hysteresis loops at 2 and 4 K for **b,** 100 nm non-templated and **e,** 100 nm templated FePc films. Inset shows an expanded version, with a coercive field of 85 and 35 mT observed at 2 K for the non-templated and templated films, respectively. No hysteresis was observed at 4 K in either of the films. **c** Schematic of the orientation of the easy magnetic plane with respect to applied magnetic field B for both the templated (top) and non-templated (bottom) films. **f** Field-dependent magnetisation of the nanowires, showing a large hysteresis of > 1 T at 2 K.

As illustrated in Figure 2d, for non-templated films the molecules lie at an angle of 82° to the substrate surface. Although all the FePc grains orient with the (100) plane parallel to the substrate, they can adopt any orientation azimuthally, *i.e.* within the plane of the film, see Figure 4c. Therefore the molecular orientation is distributed over 360° with respect to the magnetic field applied parallel to the substrate. The sharp rise at low field (up to 0.1 T) can be attributed to grains where molecules are oriented parallel to the field. As the external field increases, the spins which are not oriented favourably will be forced to align along the field away from their easy plane, leading to a slow increase in the magnetic moment. In the templated film, however, the molecules form angles of 9.0° and 7.5° with the substrate, Figure 2e and 4c. As the field is now applied almost parallel to the molecular plane the films are closer to magnetic saturation at much lower fields.

The nanowires exhibit three distinct magnetic behaviours, Figure 4f: above 60 K they behave as conventional paramagnets, with the magnetisation increasing nearly linearly with the field. Below approximately 40 K the magnetic moment increases rapidly at low fields, with a further increase in the field causing just a small change in the magnetic moment. This behaviour is similar to that observed in FePc thin films, indicative of canted ferromagnetism with the easy plane of the spins not in the same direction as the applied magnetic field. As the magnetic field increases the Zeeman energy starts to dominate, forcing the spins to rotate away from their easy plane and align along the direction of the magnetic field, leading to a slower increase of the magnetisation. As for the FePc thin films, we observe hysteresis loops below 4 K. The magnetic moment per molecule is approximately 1.5 $\mu_B$ at low fields, reaching ~ 1.8 $\mu_B$ at 7 T, although not quite reaching saturation due to the canted ferromagnetic behaviour. This is similar to the values and behaviour of our films and those reported and discussed by Evangelisti *et al*. for their bulk powders.[23] The values of the coercivities, however, are much larger with $H_c$ > 1 T at 2 K, an order of magnitude larger than the highest value currently reported for FePc thin films, and two orders of magnitude higher than for the shorter oriented wires reported previously.[25] This coercivity cannot only be due to the change in molecular orientation as identified for the films. Instead, it can be attributed to an increase in the crystal size, with the nanowires being highly crystallized for at least hundreds of nanometers along their lengths, and to the strong shape anisotropy. This can be understood both in the classical sense as observed for three-dimensional systems, as well as in terms of increased lifetime of the 1D magnetisation of the chains. The magnetisation hysteresis also displays kinks at low applied fields which can be attributed to the distribution of nanowire (and therefore magnetic domain) sizes, with the larger domains giving rise to the high coercivity contributions.

The temperature-dependent magnetisations of films and wires give more insights about the anisotropy[47] and the strength of the magnetic interactions. A Curie-Weiss fit for the films from 30 to 100 K was obtained for the inverse differential susceptibility ($\chi^{-1}$), calculated using the temperature-dependent magnetisation curves, Figure 5a. Similar high-temperature Curie-Weiss constants, $\theta_p$, were found to be 20±2 K and 27±2 K for the non-templated and templated films, respectively. These values are slightly lower than previous reported ones obtained using AC susceptibility[23] or magnetic circular dichroism (MCD).[20] However these techniques employ high frequencies or sizeable external fields, and have been shown to overestimate magnetic blocking temperatures in double decker Pcs; therefore they are expected to increase the measured $\theta_p$ in this system too.[48, 49] The Curie constants extracted from the fits between 30 and 80 K are (1.80±0.05) x $10^{-5}$ and (1.46±0.06) x $10^{-5}$ $m^3$ K $mol^{-1}$ for the non-templated and templated case, respectively. Using a total angular momentum quantum number of 1, this yields $g_{//}$ = 2.16±0.08 for the templated samples. The non-templated case has a value of $g$ = 2.40±0.06, which represents a powder average of both perpendicular and parallel contributions. This is close to the literature value of $g$=2.54 derived for powders by Evangelisti *et al*.[23] The lower value of g in the parallel direction compared to the perpendicular is in apparent conflict with XMCD results indicating that the anisotropic orbital moment is higher in the molecular plane direction.[26] However, more recent work by the same group has identified that trends in XMCD intensities are opposite to those expected from the magnetic anisotropy, indicating that the correlation of g-factor values with anisotropy in FePc is not straightforward.[45]

Increasing the crystallinity of the films and grain size has already been shown to have a large effect on the magnetic properties.[32] These effects can be further explored with the FePc nanowires.

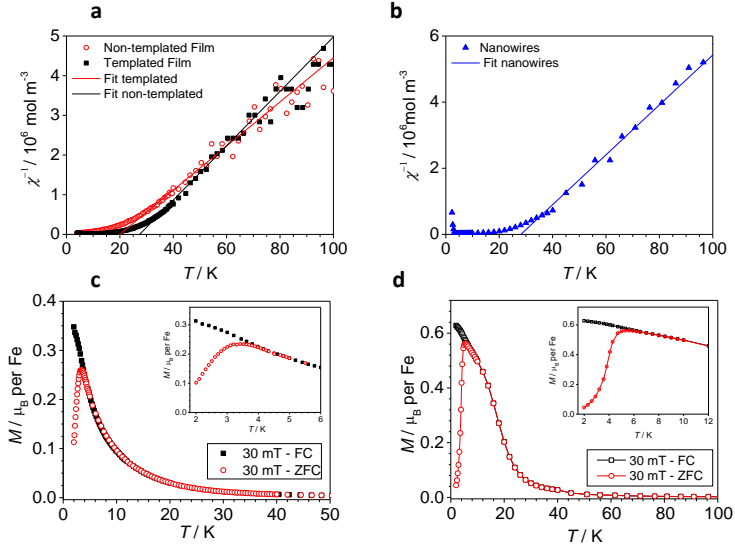

**Figure 5. Temperature dependence of magnetic properties.** **a**, Inverse differential susceptibility ($\chi^{-1}$) for FePc films. The linear fits from 30 to 100 K give Curie-Weiss constants of approximately 20 and 27 K for the non-templated and templated films, respectively. **b.** For the nanowires, the Curie-Weiss fits over the same range produces an intercept of ~ 28 K. **c.** Temperature dependent magnetisation curves measured at applied field of 30 mT using both ZFC and FC protocols for the non-templated film. The inset shows the bifurcation of magnetisation observed at approximately 4 K. Similar results are obtained for templated samples. **d**. Temperature dependent magnetisation curves measured at applied field of 30 mT using both ZFC and FC protocols for the nanowires. The inset shows the bifurcation of magnetisation observed at approximately 7 K.

When extracting the susceptibility of the FePc wires from the Curie-Weiss plot over the 30-80 K range as for the films a ferromagnetic exchange coupling constant of approximately 28±1 K, i.e. similar to the value for the templated FePc thin films, is found. The Curie constant of 1.3±0.1 x 10$^{-5}$ m$^3$ K/mol corresponds to $g$ = 2.1±0.1, which is also similar to the value for the templated films, despite a lack of preferential orientation with the molecular planes parallel to the field. The discrepancy from the value for the untemplated films could be due to the different crystal structure of the nanowires. Indeed, in addition to its strongly orientation-dependent total angular momentum,[26] FePc is characterised by a range of different low-lying magnetic crystal field states whose relative energies could be modified depending on the crystal structure.[45]

We also measured the zero-field cooled (ZFC) and field cooled (FC) magnetisation of the nanowires, Figure 5d. The magnetisation increases rapidly with decreasing temperature below approximately 30 K for both protocols. For the FC curves the magnetisation increases almost linearly with decreasing temperature but a bifurcation of magnetisation was observed at approximately 7 K in the ZFC curves for FePc nanowires and 4 K for the thin films (see Figure 5c for the non-templated samples, with similar trends observed for templated films). This bifurcation is attributed to frustrated magnetic dipole interactions between exchange coupled ferromagnetic chains or chain fragments, as previously observed for other 1D systems,[50, 51] and FePc powders.[23] As the temperature decreases, the magnetisation of each magnetic chain increases but the frustrated dipole interaction between the adjacent chains also increases. This competes with the ferromagnetic exchange interactions and leads to a decrease in the magnetic moment. During the cooling process the sample therefore experiences several ordering transitions from paramagnetism to ferromagnetism, followed by another transition to spin glass ordering. This is considered to be classic re-entrant spin glass behaviour.[52]

The values of the exchange and anisotropy energies were previously extracted for powders using a theoretical expression for the parallel susceptibility derived by O'Brien *et al.* for the one-dimensional $S = 1$ system [(CH)$_3$NH]NiCl$_3$.H$_2$O and (C$_9$H$_7$NH)NiCl$_3$.$\frac{3}{2}$H$_2$O and described by the Hamiltonian:[23, 53]

$$H_0 = -2J \sum_{i=1}^{N-1} S_{z,i} S_{z,i+1} + D \sum_{i=1}^{N} S_{z,i}^2 - g\mu_B H_z \sum_{i=1}^{N} S_{z,i} \qquad (1)$$

Notwithstanding its popularity, even for molecular systems with negligible in-plane anisotropy[54], applying such an Ising Hamiltonian to the easy-plane situation revealed by magnetometry for FePc is not valid.
In particular, assuming Hamiltonian (1) for the templated films and using $S = 1$ and $g = 2.16$ we find $D = 58\pm2$ K > $2J = 55\pm1.6$ K for the red line Figure 6a, which compares favourably with the values of $D = 53.5$ K and $J = 25.7$ K previously derived for powders using the same analysis.[23] What this solution to the Ising model however implies is that the peak at 12 K is a single-ion spin flop transition below which the moments are disordered and transverse to the z-axis, which contradicts all of the evidence for ferromagnetism in the magnetisation data.

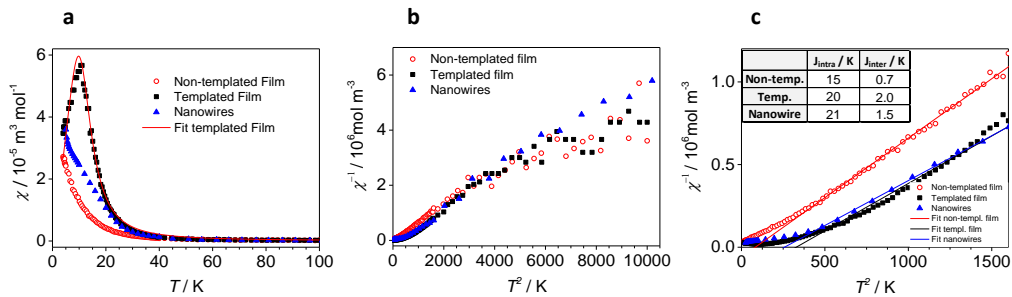

**Figure 6. Fits of the temperature-dependent magnetic properties. a.** Susceptibility for all samples, with the templated case fitted to the 1D Ising chain model (equation 1). The field was applied parallel to the substrate. **b.** Inverse susceptibility as a function of $T^2$, showing a linear behaviour up to 10000 K$^2$ **c.** Low temperature region of the $\chi^{-1}(T^2)$ curves, with the linear fit. The inset summarises the values of the intercepts, and the values of $J_{intra}$, derived from $\theta_p$, and $J_{inter}$, derived from equation 2.

Given the magnetometry data and the problems with the Ising description used previously for FePc, we introduce a more accurate description of the crystals as assemblies of 1D ferromagnetic chains with easy-plane (xy) anisotropy. Thus FePc becomes similar to the much studied inorganic compound $CsNiF_3$, composed of S = 1 ($Ni^{2+}$) chains with intrachain ferromagnetic correlations and an easy-plane anisotropy.[50] The magnetic behaviour has been studied by many workers, with for example numerical solutions provided by Loveluck et al.[55] and quantum corrections introduced by de Neef.[56] It was noted that when the field is applied perpendicular to the chain axis, the susceptibility, $\chi_{chain}$, within an isolated chain with magnetic interactions $J_{intra}$, diverges with temperature T according to the power law $2J_{intra}T^{-2}$,[51] precisely what is predicted for the one-dimensional xy ferromagnet.[57] The standard mean field approximation, where chains impose self-consistent fields on each other, implies that an assembly of chains, weakly coupled to each other with an interaction $J_{inter}$ will follow a modified Curie-Weiss law where the susceptibility of the 3D crystals, $\chi_{assembly}$, is described by:

$$\chi_{assembly} = \chi_{chain}/(1-\chi_{chain} z' J_{inter}) = 2 J_{intra}/(T^2 - 2z' J_{intra} J_{inter}) \qquad (2)$$

where z' is the number of neighbouring (interacting) chains, which in our case is 4. This suggests plotting $1/\chi$ against $T^2$ for the nanowires and granular films. Fig. 6b shows the results, and quite remarkably, for all samples, the modified law is satisfied over most of the $T^2$ range measured. The intercept (measuring the product $2z'J_{intra}.J_{inter}$) has a value of 258±15 $K^2$ for the wires while being substantially smaller at 83±8 $K^2$ for the untemplated nanograins, and larger, at 328±20 $K^2$ for the templated films, Figure 6.c. The simple picture and associated formula, where there is a positive intercept for $1/\chi$ against $T^2$, naturally accounts for the onset of bulk ferromagnetism as evidenced by the magnetisation sweeps at temperatures close to that defined by the intercept. The intrachain couplings can be deduced from the *high temperature* Curie-Weiss constants using

$$J_{intra} = 3\theta_p/zS(S+1) \qquad (3)$$

where z, representing the nearest neighbours within a chain, is 2, and the results summarised in Table 1. The interchain couplings $J_{inter}$ are extracted from the intercepts and the Curie-Weiss-derived $J_{intra}$, and are dependent on the orientation of the easy-plane with respect to the field, with the value for the templated films ($J_{inter}$ = 2.0±0.2 K) substantially larger than for the untemplated films ($J_{inter}$ = 0.7±0.1 K). This could be due to the larger intrachain couplings, and higher low-field magnetisation, but also due to the brickstack arrangement of the molecules with parallel molecular planes between columns, rather than the herringbone structure that has been observed in the non-templated films, where two inequivalent orientations exist between columns.[30] The nanowires present an intermediate value $J_{inter}$ = 1.5±0.1 K. The wires also contain herringbone arrays with similar interchain displacements as in the non-templated films; the higher value compared to the non-templated grains is likely due to the much longer chain domains in the wires compared to the films.

Figure 7 shows $\chi$ as a function of both temperature and magnetic field. For all polymorphs, $\chi(H, T)$ is characterised by two major ridges, the first emanating from the origin along the T = 0 axis and the second extending from the H = 0 axis towards large H and T. The former is dependent on the detail of the interchain interactions as well as the competition between dipolar interactions and terms accounting for the microscopic interactions in equation 2, and we will not consider it further here. The extensive literature on CsNiF$_3$ and related materials[50, 58, 59] suggests that magnetic solitons, which are mobile domain walls, make a major contribution to the latter which dominates the one-dimensional regime at higher *T*.

Because the soliton is a domain wall, the spins within it experience less exchange energy than those in the quasi-ordered centres of the domains, implying that they will dominate the response to external fields. Therefore $\chi$ is largely sensitive to changes in thermally activated soliton populations with magnetic fields, from which it follows that the ridges where $\chi$ is maximum for fixed *H* occur roughly at temperatures corresponding to the soliton energies *E*. Figure 7a shows schematically the solitons that we expect to contribute.

The Hamiltonian for the easy-plane ferromagnetic chain is:

$$H_0 = -J \sum_{i=1}^{N-1} S_i S_{i+1} + D \sum_{i=1}^{N} S_{z,i}^2 - g\mu_B H . \sum_{i=1}^{N} S_i \tag{4}$$

Depending on the field orientation and strength, the solitons take on different appearances. Figure 7a illustrates the limiting cases of a moderate field (i.e. $|H| \sim D$ and we also assume $D \sim J$) parallel (on the left in the panel) and perpendicular (on the right) to the easy plane. The energy for a soliton of the former type, corresponding to the situation of the templated films, can be calculated using for example the formalism derived by Samalam and Kumar[60]:

$$E(T,H) = E_0 \left\{ 1 - \frac{kT}{4\sqrt{DJ}} \left[ \frac{1}{\sqrt{2}} + \frac{1}{4}\sqrt{\frac{g\mu_B H}{D}} \right] \right\} \tag{5}$$

where $E_0 = 8\sqrt{Jg\mu_B H}$.

The ridge in the (H,T) plane is then where the condition $E(T,H) = k_B T$ is met. The best description of the experimentally observed ridge for the templated films (Figure 7b) is obtained for *D* = 33±3 K when we constrain *J* = 20 K and g = 2.16, the values obtained from the Curie-Weiss plots. Note that the computed ridge takes the form of a parabola.

For the non-templated films, the ridge in the $\chi$(*H*, *T*) map (Figure 7c) appears more linear and somewhat sharper than for the templated films, differences which arise because many molecular chains are oriented such that the (x,y) plane is not parallel to the field and for which equation (4) is therefore not valid. For an easy-plane anisotropy as in the Hamiltonian (4), the susceptibility $\chi(H_x, H_y, H_z)$ is actually a function $\chi_o(H_\parallel, H_\perp)$ of only the two variables defined by the field $H_\perp = H_z$ along the hard (chain) axis and the field $H_\parallel$ with magnitude $\sqrt{H_x^2 + H_y^2}$ in the easy plane. Each chain lies essentially parallel to the substrate, and is characterised by an angle $\varphi$ relative to the external field, implying that the total susceptibility will be the average:

$$\chi_{tot}(H) = \int \chi_o(H\sin\varphi, H\cos\varphi) d\varphi/(2\pi) \tag{6}$$

For a field $H_z = Hcos\varphi \neq 0$ it is energetically advantageous for the spins in the soliton to twist towards the chain axis.

A full calculation of equation (6) is an elaborate exercise in computational physics, and because we are simply trying to understand the nature of the ridges seen in Figures 7c and d, and in particular the reason for the differences from their appearance in Figure 7b, we now proceed phenomenologically and focus on what happens when there is a strong field component $H_z$ along z. What we can stabilize here is the Neel soliton labelled "non-templated" (as opposed to the Bloch soliton labelled "templated") in Figure 7a, where the spin rotates through the hard axis. The soliton carries an effective Ising spin degree of freedom, pointing along the hard axis but with the possibility to lie parallel or antiparallel to the field along z, and whose extent is proportional to the healing length $\sqrt{\frac{J}{2D}}$ for twists of the magnetisation towards the hard (perpendicular to the molecules) magnetic axes. Its energy is therefore the sum of soliton creation and Zeeman terms:

$$E = \sqrt{8DJ} \pm g\mu_B H\pi \sqrt{\frac{J}{2D}} \qquad (7)$$

We consider the high energy branch which describes the upper edge of the ridge as *H* and *T* both increase, and correct the zero-field wall formation energy for thermal disorder to first order in *T* by adding a linear term for the temperature dependence. We then obtain:

$$E = \sqrt{8DJ} - \frac{T}{\alpha} + g\mu_B H\pi \sqrt{\frac{J}{2D}} \qquad (8)$$

It follows that the ridge in the (*H,T*) plots can be described by the general equation:

$$T = \alpha\sqrt{8DJ} + \beta g\mu_B H\pi \sqrt{\frac{J}{2D}} \qquad (9)$$

where *α* and *β* represent renormalization factors of order unity for the soliton creation energy and width respectively. Those values can be determined from the non-templated film, using the anisotropy derived in the highly oriented templated system, i.e. *D* = 33 K, and which should not vary as the structure within the chains is identical. Using the exchange and g-factor obtained from the Curie-Weiss fit (*J* = 15 K and *g* = 2.4), we find *α* = 0.44 and *β* = 1.52, and the corresponding soliton ridge (equation (9)) is plotted in Figure 7c. These phenomenological values for *α* and *β* can then be used to together with the nanowire Curie-Weiss values (*J*= 21 K and *g*= 2.1) to fit the nanowire ridge and thereby determine their anisotropy energy. The result, for which the fit is plotted in Figure 7d, is *D* = 62±9 K. Hence the anisotropy for the η phase wires is higher than for the α-phase films, which can be ascribed to the different structure. Note that the value of *D* can be easily extracted visually from the intercept of the fit at the *H*=0 line which corresponds to $\alpha\sqrt{8DJ}$.

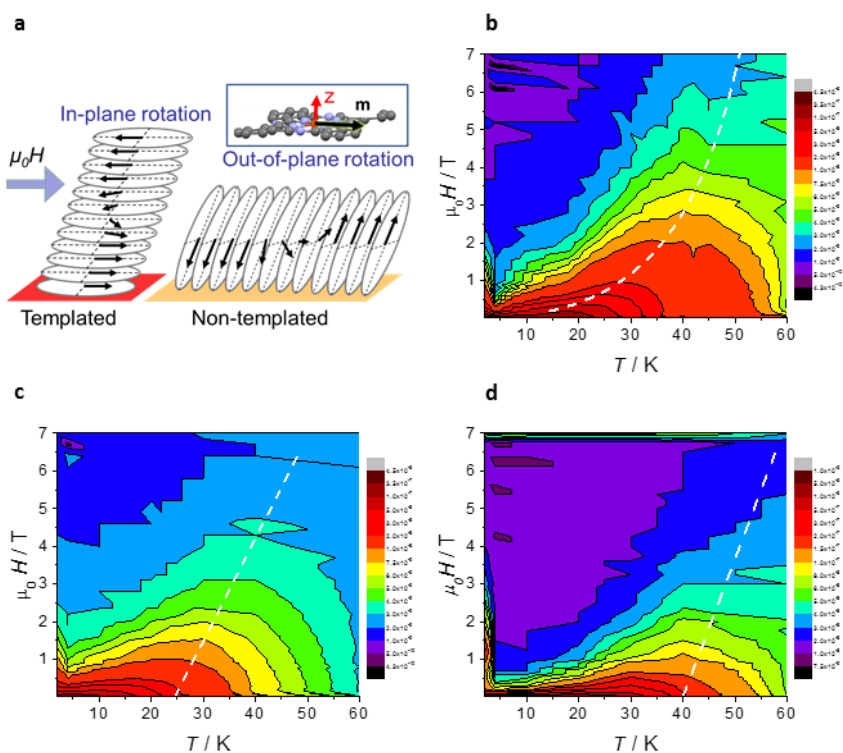

**Figure 7. Susceptibility as a function of applied field and temperature. a**. Schematic of solitons in templated and non-templated films (in the latter, representing grains whose molecular planes are normal to the applied field *H*). The ellipses represent the molecular plane, while molecular magnetic moments m are depicted by the black arrows. **b.** Susceptibility derived from the *M*(*H*) curves, plotted as a function of applied field and temperature for templated films, **c.** non-templated films and **d.** nanowires. The dashed white line are fits to equation 4 (b) or equation 7 (c-d).

To rationalise the correlation between the structure and the magnetic properties and attempt to derive design rules, we examine the intermolecular displacements within a molecular chain, using values for the isomorphous CuPc crystals as before, summarised in Figure 8. We add the β-phase crystals, which have previously been shown to have a negligible *J* and very high anisotropy of *D* = 98 K.[61] Clearly the lateral displacement of the Fe atom with respect to the neighbouring bridging N is key in determining the strength of the magnetic anisotropy, and the extremely high values for the β phase can be attributed to a nearly aligned arrangement of the Fe with a bridging N of the neighbouring molecule. It is more difficult to identify a single parameter that determines the strength of the magnetic interaction *J* and coercivity, which is increased for the nanowires. It is highly plausible that the strength of the magnetic exchange is affected by the Fe-Fe distances, which are smallest in the nanowire case. The small difference with the α-phase does however suggest that size effects might be dominating the coercivity, as already noted above.

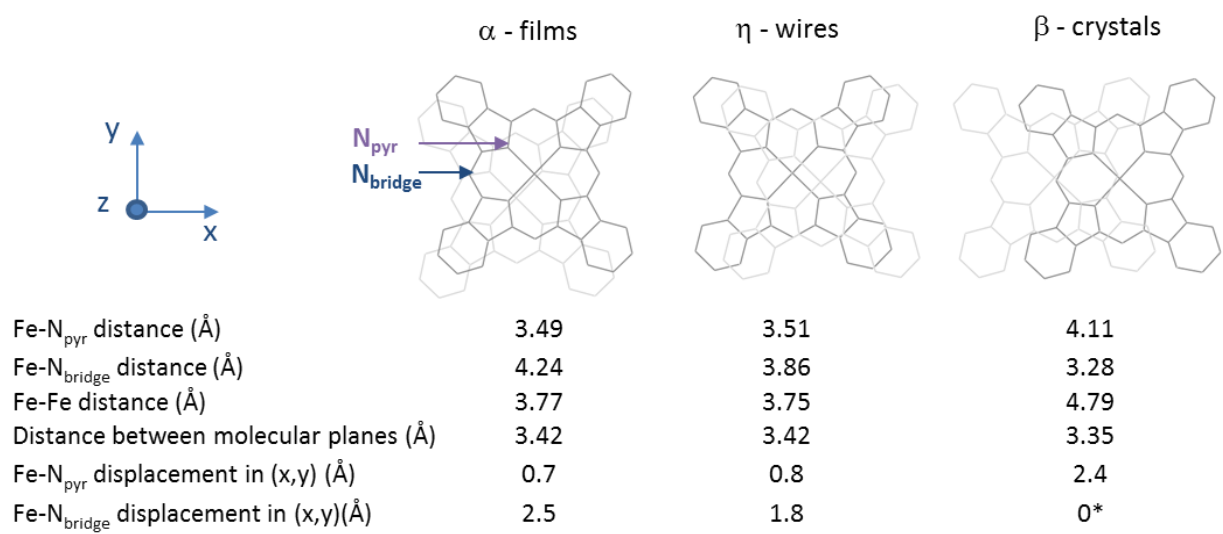

| | α - films | η - wires | β - crystals |
|---|---|---|---|
| Fe-$N_{pyr}$ distance (Å) | 3.49 | 3.51 | 4.11 |
| Fe-$N_{bridge}$ distance (Å) | 4.24 | 3.86 | 3.28 |
| Fe-Fe distance (Å) | 3.77 | 3.75 | 4.79 |
| Distance between molecular planes (Å) | 3.42 | 3.42 | 3.35 |
| Fe-$N_{pyr}$ displacement in (x,y) (Å) | 0.7 | 0.8 | 2.4 |
| Fe-$N_{bridge}$ displacement in (x,y) (Å) | 2.5 | 1.8 | 0* |

**Figure 8. Projection of two consecutive molecules in a chain along the molecular planes**, and summary of key nearest-neighbour distances between nearest equivalent atoms on two different molecules.[31, 34, 62] $N_{pyr}$ and $N_{bridge}$ are the pyrrole and the bridging nitrogens respectively. The displacements in (x,y) are calculated with reference to the average molecular plane and assuming that all atoms within a molecule are coplanar. The error on the lateral displacements is large due to atomic displacements along z, with * estimated visually.

*Comparison of the FePc polymorphs*

To highlight the outstanding magnetic properties of the FePc polymorphs we present the spin state, Curie-Weiss constant, and exchange coupling *J* of metal phthalocyanine thin films and nanowires in Table 1. Indeed the α- and η-polymorphs of FePc exhibit the highest Curie-Weiss temperatures and ferromagnetic coupling strengths amongst metal phthalocyanine films and nanowires investigated so far.

**Table 1. Summary of the main properties of key transition metal Pcs.** Comparison of experimentally determined magnetic properties of Pcs in film (60-200 nm on kapton substrates) and powder form. Curie-Weiss constants $\theta_p$ and exchange interactions $J$ were obtained from SQUID magnetometry, except for the value identified with * or †, which were obtained from MCD or AC susceptibility respectively. The values of $J$ were extracted from $\theta_p$ using equation (3), validated via its use for CuPc where the exchange constants obtained are consistent with the values marked by ** obtained from detailed (exact) 1D analysis. We do not report $\theta_p$ values for CoPc as the strong T-independent van Vleck paramagnetism prevents straightforward Curie-Weiss fitting[21] for this material with its high antiferromagnetic exchange constant, meaning that only Bonner-Fisher (exact) analysis including background terms could be used.

| Polymorph | S | Sample | $\theta_p$ (K) | $J/k_B$ (K) | References |
|---|---|---|---|---|---|
| α-MnPc | 3/2 | Film | -1.6 to -2 | - | [18, 19] |
| β-MnPc |  | Film | 11.5 | - | [63] |
| β-MnPc |  | Powder | 10 | - | [19] |
| α-FePc | 1 | Film | 20 | 15- | This work |
| α-FePc |  | Templated film | 27 | 20 or 19‡ | This work, [20] |
| α-FePc |  | Powder | 40† | 25† | [23] |
| β-FePc |  | Powder | 9 | - | [64] |
| η-FePc |  | Nanowire | 28 | 21 | This work |
| α-CoPc | 1/2 | Film | - | -107 | [21] |
| α-CoPc |  | Templated film | - | -80 | [21] |
| α-CoPc |  | Powder | - | -78 | [21] |
| β-CoPc |  | Powder | -3.75 | -1.9 | [21] |
| α-CuPc | 1/2 | Film | -2.6 | -1.4** | [19] |
| α-CuPc |  | Templated film | -3.0 | -1.6** | [19] |
| β-CuPc |  | Film | 0.0 | 0.0** | [19] |
| β-CuPc |  | Powder | -0.4 | 0.15** | [19] |
| η-CuPc |  | Nanowire | - | -1.8 | [31] |

**Conclusions**

We have prepared flexible molecular ferromagnetic films of α-FePc displaying well-controlled structural, spectroscopic and magnetic properties using molecular templating with PTCDA. This simple technique is compatible with many substrates, including transparent flexible plastics and allows for modification of the molecular orientation from nearly perpendicular to nearly parallel to the substrate. The α-phase FePc films consist of 1D ferromagnetic chains at low temperatures, and the SQUID magnetometry shows that templating modifies the direction of the magnetic easy plane. The alignment of the easy plane with the applied field and the preponderance of the brickstack interchain arrangement following templating leads to an increase in the Curie-Weiss constant, from 20 K to 27 K, a decrease in the coercive field at 2 K, from 80 mT to 35 mT, and an increase in the interchain interactions from 0.7 K to 2.0 K, as compared to the non-templated films. As charge carriers preferentially hop between the molecules within the molecular chains, the ability to control the molecular orientation through templating is not only useful for changing the direction of the ferromagnetic chains but also changes the direction with the highest charge carrier mobility with respect to the substrate, an effect that has been previously exploited for improving solar cell efficiency[30, 65, 66] We therefore have a recipe for suitably orienting ferromagnetic chains and molecules for magneto-optoelectronic devices.

Organic vapour phase deposition was then used to fabricate η-polymorph FePc nanowires. SQUID magnetometry shows that the nanowires exhibit bulk ferromagnetism below 4 K, with a Curie-Weiss constant of 28 K. A surprisingly large coercivity of over 1 T was observed at 2 K, arising from the long chain length and strong shape, and molecular, anisotropy. A bifurcation of the magnetisation between ZFC and FC temperature-dependent magnetisation curves occurred at 7 K due to frustrated dipolar interactions between the 1D ferromagnetic chains, as observed for the films, although in that case the smaller domains lowered the bifurcation to 4 K.

The key qualitative result of our experiments is that the FePc is an excellent molecular analogue of the classic magnetic soliton host $CsNiF_3$, containing chains of ions with xy spins easily rotating in the plane perpendicular to the chains. The xy nature of FePc leads to both the dominance by solitons of the field- and temperature-dependent susceptibility when H or T are well away from zero, and the super-Curie Weiss divergence $1/(T^2-\theta)$ (a form not involving the exponential $\exp(J/T)$ as for Ising magnets) of the zero field susceptibility at low to medium temperatures. Developing formalisms for Bloch and Neel solitons, we derive a magnetic anisotropy of $D$ = 33 K for the films, and 62 K for the wires. The high anisotropy in the wires correlates with a decrease in the displacement between the Fe atom and the bridging N in the neighbouring molecule compared to the α-phase films. What of course differentiates $CsNiF_3$ from FePc is that the latter is a semiconductor, for which (in untemplated film form) we have already demonstrated transistor action.[22]

Our summary of key magnetic properties for phthalocyanine molecules in a range of forms including films, nanostructures and powders, Table 1, highlights that FePc stands out as a ferromagnetic system with favourable properties for investigations of magneto-optical and -electrical effects, as well as quantum ferromagnetism.[67, 68] We especially look forward to magnetotransport experiments exploiting the methods of ref. [22] which describes the excellent performance of simply fabricated TMPc nanowire transistors.

**Acknowledgements**

The authors wish to acknowledge R. Sweeney, M. Ardakani and M. Ellerby for help with XRD, TEM and SQUID, respectively. We thank Dr Wei Wu and Professors Andrew Fisher and Nicholas Harrison for useful discussions. Financial support from the Engineering and Physical Sciences Research Council (EPSRC) (EP/F039948/1 and Basic Technology grant "Molecular Spintronics" EP/F04139X/1, EP/F041160/1) and the National Natural Science Foundation of China (61704017) is gratefully acknowledged. We thank Professor Thomas Anthopoulos for the provision of the tube furnace used in the OVPD experiments. P. R. and S. H. thank Dr Salahud Din and Kurt J. Lesker Company for technical support and funding. P. R. thanks the department of Materials at Imperial College for the provision of a PhD studentship.